\documentclass[trackchanges]{aastex701}

\begin{document}

\title{The role of specific entropy in the onset of the luminosity bump}

\author[orcid=0000-0002-1463-726X,sname='Hekker']{S. Hekker}
\altaffiliation{Heidelberg Institute for Theoretical Studies (HITS) gGmbH, Schloss-Wolfsbrunnenweg 35, 69118 Heidelberg, Germany}
\affiliation{Heidelberg Institute for Theoretical Studies (HITS) gGmbH, Schloss-Wolfsbrunnenweg 35, 69118 Heidelberg, Germany}
\email[show]{saskia.hekker@h-its.org}  
\altaffiliation{Landessternwarte K{\"o}nigstuhl (LSW), Heidelberg University, K{\"o}nigstuhl 12, 69117 Heidelberg, Germany}
\affiliation{Landessternwarte K{\"o}nigstuhl (LSW), Heidelberg University, K{\"o}nigstuhl 12, 69117 Heidelberg, Germany}

\begin{abstract}

During their evolution, stars follow a distinct path in luminosity--temperature space. Low-mass stars that have exhausted hydrogen in their core follow the so-called red-giant branch which is predominantly in the direction of increasing luminosity and decreasing surface temperature. The luminosity bump is a temporary decrease in luminosity against this otherwise increasing trend. The bump is present in both observations and stellar evolution models. However, with canonical physics included in the models, the bump in computed evolutionary tracks appears at higher luminosities than the observed bump. To understand why the bump appears later in the models than in observations, the physics of the bump needs to be unravelled. The end point of the bump is well-understood; however, the onset of the bump is still an enigma. Here, we report on the physical origin of the onset of the luminosity bump. We show that the difference in specific entropy at the mean molecular weight discontinuity decreases due to the discontinuity moving in. This decrease is attributed to the decrease in the ratio of the temperature to the pressure at smaller radii. Upon reaching a critical value the specific entropy difference at the mean molecular weight discontinuity is reduced sufficiently to reduce the specific entropy in the convective envelope. The latter is a key signature of the bump. Hence the evolution of the specific entropy at the mean molecular weight discontinuity provides a viable description for the onset of the bump. 

\end{abstract}

\keywords{\uat{Stellar astronomy}{1583} --- \uat{Stellar evolution}{1599} --- \uat{Red giant stars}{1372} --- \uat{Stellar structures}{1631}}


\section{Introduction} 

Red-giant branch (RGB) stars of masses below roughly 2~M$_{\odot}$ experience the red-giant-branch bump (RGBB) which is visible as a zig-zag in the evolutionary track as for example shown by its evolution of luminosity ($L$) versus effective temperature ($T_{\rm{eff}}$) (see Fig.~\ref{Fig:HRD}). This is a well-known feature in stellar models, and is also clearly visible in observations of, for instance, open and globular clusters. The phenomenon causes a star to live longer in a narrow band of luminosities and hence, in an iso-age population, an increased stellar density is observed at the luminosity of the bump. Consequently, the RGBB serves as an important reference point to calibrate models \citep[e.g.][and references therein]{riello2003,angelou2015,joyce2015,khan2018}. A major point of attention is the fact that the bump in stellar evolution tracks including canonical physics does appear at a higher luminosity compared to observations \citep{khan2018}. To mitigate this, it is common practise to change the amount of overshoot at the base of the convection zone, however this creates tension with the Li surface abundance \citep{cao2026}. Hence, additional changes to the current physics included in the models may be necessary. To infer what changes may be required, we investigate here the physics of the bump. \citet[][and references thererin]{jcd2015} did look into the physics of the onset of the bump via homology relations. They do however find substantial differences between their expressions and the models. They attributed this to the rather drastic approximations they made in the derivations of the equations. Though they follow \citet{refsdal1970} in concluding that the cause of the luminosity variation in the bump is predominantly the result of the
effect of the composition discontinuity on the hydrostatic
equilibrium. This was indeed confirmed by \citet{hekker2020} showing that stellar evolution tracks of models without the composition discontinuity do not experience the RGBB. Here we take the next step, and we further investigate the role of the composition discontinuity in the onset of the bump. To do so, we will first provide a description of the structure of stars on the RGB in the vicinity of the bump.

\subsection{Stellar structure of RGB stars}
RGB stars have an inert helium core surrounded by a hydrogen-burning shell, both located in a region where energy transport takes place via radiation. Above this radiative region, these stars exhibit a convective outer envelope \citep[e.g.][]{kippenhahn2013}. While ascending the RGB, stars have an expanding surface and a contracting core. This causes the deepest layer of the convective envelope to move inwards reaching layers that have changed their chemical composition due to (burning) processes earlier in the evolution of the star. Incorporating deeper radiative layers into the convective region changes the chemical mixture. This is referred to as dredge-up. This change in composition is seen in surface layers, due to the (near) instantaneous mixing taking place in convective conditions \citep[e.g.][]{sneden1991,gratton2000,roberts2024}. Deep in the star, the dredge-up creates a mean molecular weight discontinuity ($\mu$-disc) at the base of the convection zone (BCZ, see Fig.~\ref{Fig:absprofiles}). The difference in mean molecular weight at this discontinuity ($\delta\mu$) increases as long as the BCZ moves to deeper layers. This increase stops when the BCZ reaches the contracting part of the star and starts retracing its route back towards the surface of the star, referred to as the receding of the BCZ. The $\mu$-disc is now located in the radiative part of the star which means $\delta\mu$ remains largely unchanged due to the lack of an efficient mixing process. Additionally, the $\mu$-disc is now located in the contracting part of the star. This means that the $\mu$-disc moves to smaller radii while remaining at the same mass coordinate upon further evolution \citep{hekker2020}.

As mentioned above, the $\mu$-disc is known to be an essential ingredient for the bump \citep[e.g.][and references therein]{kippenhahn2013,jcd2015,hekker2020}, but until now it was unclear how. A common hypothesis is that the bump is a response to the hydrogen-burning shell burning through the $\mu$-disc \citep{kippenhahn2013}. As shown by \citet{jcd2015} this explains the end of the temporary decrease in luminosity at $L_{\textrm{\tiny{min}}}$ (see Fig.~\ref{Fig:HRD}), which marks the endpoint of the bump. It does, however, not explain the onset of the decrease in luminosity at $L_{\textrm{\tiny{max}}}$. Here we provide an explanation for this onset, i.e., for the start of the bump.

\section{Specific Entropy} 

\begin{figure}
\centering
\includegraphics[width=\textwidth,clip]{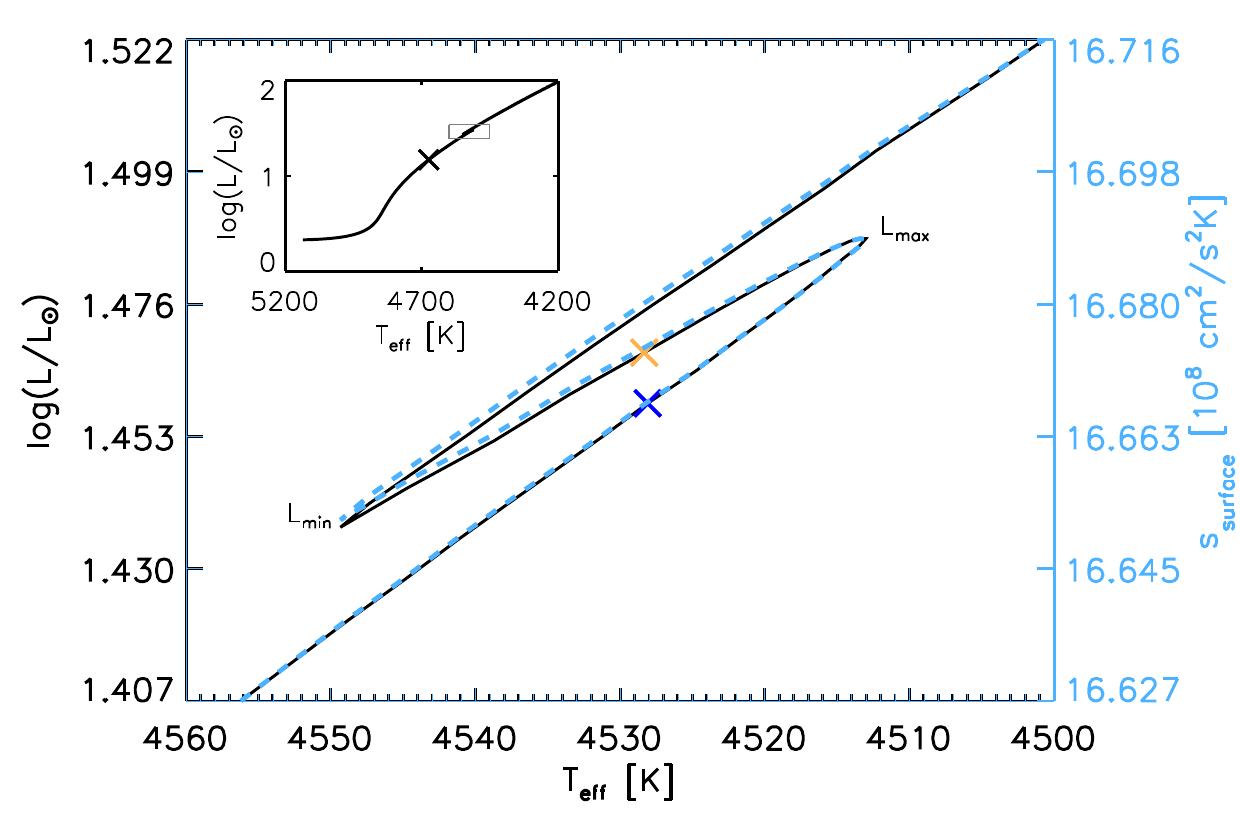}
\caption{Inset: Hertzsprung-Russell diagram showing a 1~M$_{\odot}$ track from the subgiant branch till the tip of the RGB. The general direction of evolution is from low to high luminosity. The stellar evolution track is the same as used by \citet{hekker2020}. The bump is highlighted by the grey box. Main figure: a close-up of the bump with the luminosity maximum and minimum ($L_{\textrm{\tiny{max}}}$  and $L_{\textrm{\tiny{min}}}$) indicated. The specific entropy at the surface is shown by the dashed light blue line and the axis on the right. The three crosses indicate the location of the models shown in Fig.~\ref{Fig:absprofiles}.\label{Fig:HRD}}
\end{figure}

\begin{figure}
\centering
\includegraphics[width=\columnwidth]{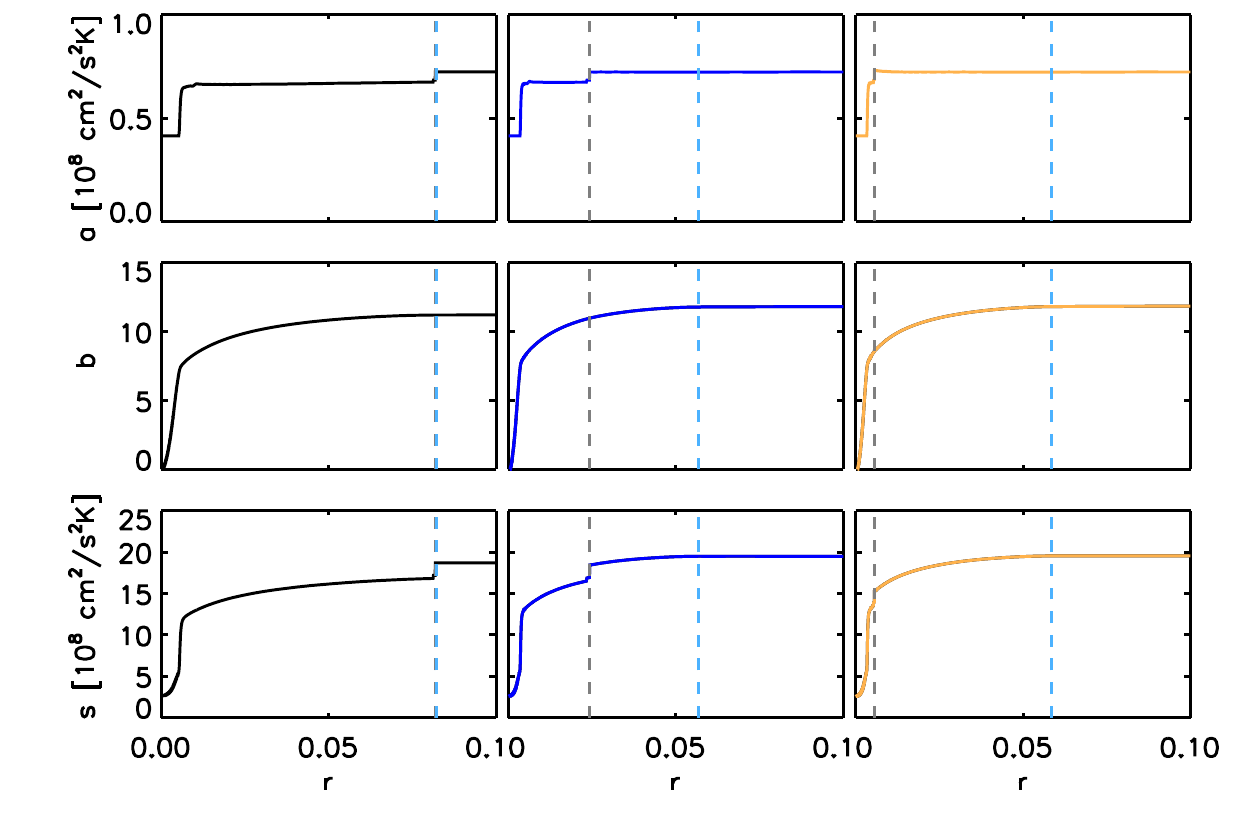}
    \caption{The values for $a$, $b$, and $s$ (top to bottom) as a function of fractional radius ($r$) for the models indicated with the black, blue and orange crosses in Fig.~\ref{Fig:HRD} (left to right). Only the inner 10\% in radius of the stellar models is shown here, as this is the region of interest. The vertical grey and light blue dashed lines indicate the location of the $\mu$-disc and BCZ, respectively. Note that for the model in the left column, the $\mu$-disc and BCZ are co-located, and the vertical lines overlap.\label{Fig:absprofiles}}
\end{figure}

The specific entropy ($s$) is the quantity we will use to study the onset of the bump. This quantity is of use for two reasons. 
First, it is directly related to the luminosity ($L$). According to the Stefan-Boltzmann law, $L$ is given by:
\begin{equation}
    L=4\pi \sigma R^2 T_{\textrm{\tiny{eff}}}^4, 
    \label{Eq:L}
\end{equation}
    where $T_{\textrm{\tiny{eff}}}$ is the effective temperature at the stellar surface, $R$ the stellar radius and $\sigma$ the Stefan-Boltzmann constant \citep{kippenhahn2013}. Along the RGB the surface temperature decreases while the luminosity increases, which indicates that the star expands. This expansion is related to the specific entropy through $\epsilon_{\textrm{\tiny{g}}}= -T\partial{s}/\partial{t}$ with $T$ the local temperature and $\partial{}/\partial{t}$ indicating a time derivative. The sign of $\epsilon_{\textrm{\tiny{g}}}$ indicates whether the star expands ($\epsilon_{\textrm{\tiny{g}}}<0$) or contracts ($\epsilon_{\textrm{\tiny{g}}}>0$). Hence, an increase in $s$ with time at the surface indicates an expanding star, i.e., an increase in $R$, leading to an increase in luminosity and a decrease in surface temperature, and vice versa (see Fig.~\ref{Fig:HRD}). This is substantiated by the fact that the stars on the RGB evolve on a path closely resembling a Hayashi-track where the temperature is a function of the luminosity. This immediately implies luminosity to be a function of $R$ and thus of the specific entropy.

Second, for an ideal gas\footnote{The assumption of an ideal gas is valid in a large part of the star.} the specific entropy is related to the mean molecular weight \citep{hansen1994}:
\begin{equation}
     s=\frac{N_{\textrm{\tiny{A}}}k_{\textrm{\tiny{B}}}}{\mu}\ln[(T/T_0)^{5/2}/(P/P_0)]+c,
     \label{Eq:s1}
\end{equation}
with $N_{\textrm{\tiny{A}}}$ Avogadro’s constant, $k_{\textrm{\tiny{B}}}$ the Boltzmann constant, $\mu$ mean molecular weight, $P$ pressure, with the subscript $0$ indicating reference values, and $c$ is an integration constant. To separate out the dependence of $s$ on the mean molecular weight from the dependence on the temperature and pressure, we rewrite Eq.~\ref{Eq:s1} to $s=ab+c$ with $a= N_{\textrm{\tiny{A}}} k_{\textrm{\tiny{B}}}/\mu$ and $b=\ln[(T/T_0)^{5/2}/(P/P_0)]$. Profiles for $a$, $b$ and $s$ as a function of fractional stellar radius are shown in Fig.~\ref{Fig:absprofiles} for the models indicated with crosses in Fig.~\ref{Fig:HRD}. These profiles show that $b$ decreases towards the stellar centre, while $a$ is very similar between all three models except for the location of the $\mu$-disc.

\subsection{Specific entropy as a function of stellar radius}
The specific entropy increases with radius in the deep radiative interior up until the BCZ. In the convection zone the specific entropy remains constant\footnote{This is true except for the super-adiabatic layer close to the surface.} \citep[see bottom panels in Fig.~\ref{Fig:absprofiles} as well as Eq. 3 and associated text in][]{hekker2020}. We therefore investigate the specific entropy at the BCZ ($s_{\textrm{\tiny{BCZ}}}$) as this is representative of the behaviour of the specific entropy at the stellar surface. To gain insight in the contribution of the $\mu$-disc to the specific entropy at the BCZ, we divide the specific entropy into different contributions, and consider the specific entropy below, at and above the radius of the $\mu$-disc separately:
\begin{equation}
    s_{\textrm{\tiny{BCZ}}}=  s(r_0)+ \Delta s_{\textrm{\tiny{below}}} + \delta s+ \Delta s_{\textrm{\tiny{above}}},
    \label{Eq:s2}
\end{equation}
where $s(r_0)$ is the specific entropy at the centre of the star, $\Delta s_{\textrm{\tiny{below}}} =\lim_{r \nearrow r_{\mu \textrm{\tiny{-disc}}}} s(r)-s(r_0)$ the increase in specific entropy in the region below the $\mu$-disc, $\delta s$ the difference in specific entropy at the $\mu$-disc, and $\Delta s_{\textrm{\tiny{above}}} =s(r_{\textrm{\tiny{BCZ}}})-\lim_{r \searrow r_{\mu \textrm{\tiny{-disc}}}} s(r)$ the increase in specific entropy in the region above the $\mu$-disc and below the BCZ. 

\subsection{Specific entropy as a function of time}
\begin{figure}
\centering
\includegraphics[width=\columnwidth, clip]{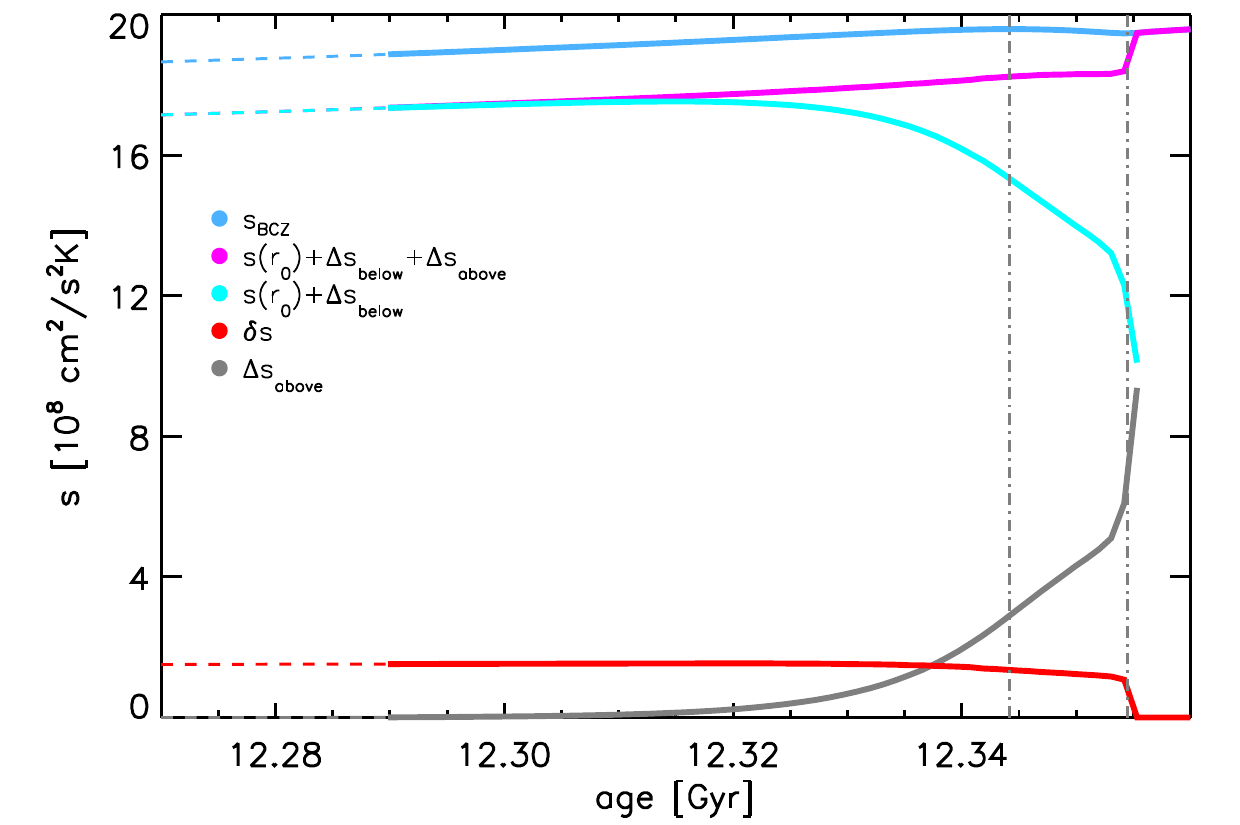}
\caption{Specific entropy at the BCZ ($s_{\textrm{\tiny{BCZ}}}$) indicated in blue (this is a representative proxy of the specific entropy at the surface ($s_{\textrm{\tiny{surface}}}$) shown by the blue curve in Fig.~\ref{Fig:HRD}) as a function of time. The different contributions to $s_{\textrm{\tiny{BCZ}}}$ as described by Eq.~\ref{Eq:s2} are indicated as per the legend: the contributions below, at and above the $\mu$-disc are shown in cyan, red and grey, which add up to $s_{\textrm{\tiny{BCZ}}}$. The magenta curve shows the total entropy in the radiative layers excluding the contribution at the $\mu$-disc. All quantities are shown as solid lines when the BCZ is receding. The vertical grey dashed-dotted lines indicate the time of luminosity maximum and minimum.\label{Fig:sparts}}
\end{figure}

\begin{figure}
\centering\includegraphics[width=\columnwidth, clip]{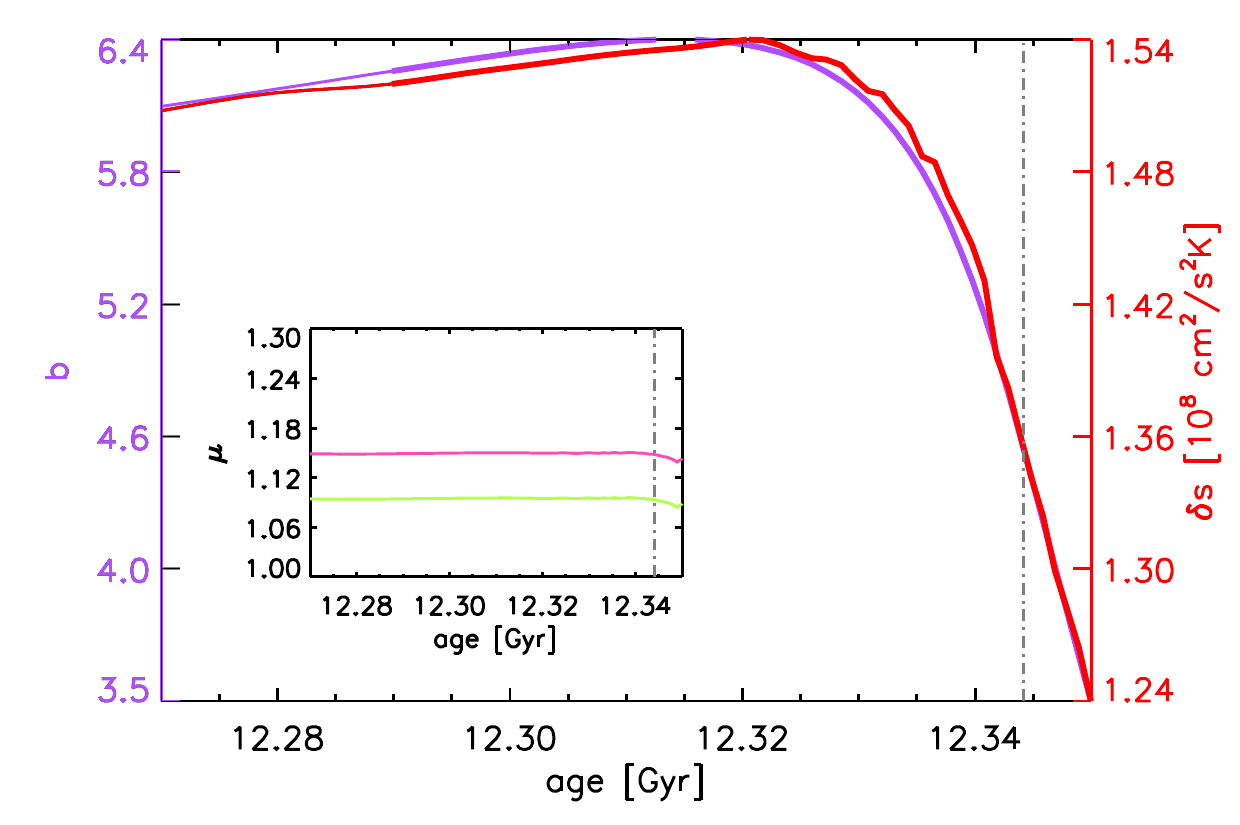}
\caption{The evolution of $b$ (purple), and the step in specific entropy ($\delta s$, red) both at the $\mu$-disc. The Pearson correlation coefficient between these curves is 0.997. The inset shows the (constant) mean molecular weight just above (green) and below (pink) the discontinuity. The vertical dashed-dotted line indicates the luminosity maximum of the bump.\label{Fig:bds}}
\end{figure}

Following the adoption of Eq.~\ref{Eq:s2} to describe the specific entropy as a function of radius, we now look at changes in specific entropy over time. When stars ascend the RGB, the specific entropy in the envelope, i.e. $s_{\textrm{\tiny{BCZ}}}$, increases. This is the case on the entire RGB, except between the luminosity maximum and minimum of the bump \citep[$L_{\textrm{\tiny{max}}}$ and $L_{\textrm{\tiny{min}}}$, see Fig.\ref{Fig:HRD} and][]{hekker2020}.
We focus on the relatively short time interval starting just before the luminosity maximum of the bump till the luminosity minimum of the bump: as long as the $\mu$-disc coincides with the BCZ, $\Delta s_{\textrm{\tiny{above}}}=0$. When the BCZ starts receding $\Delta s_{\textrm{\tiny{above}}} >0$, while at the same time $\Delta s_{\textrm{\tiny{below}}}$ becomes smaller (see Fig.~\ref{Fig:sparts} for the evolution of $s_{\textrm{\tiny{BCZ}}}$ and the different terms specified in Eq.~\ref{Eq:s2}). In Fig.~\ref{Fig:sderivs}, we show the time derivatives of the different contributions to $s_{\textrm{\tiny{BCZ}}}$. 
After the time at which the BCZ starts receding, the $\mu$-disc gradually moves to smaller radii, i.e., $r_{\mu \textrm{\tiny{-disc}}}$ becomes smaller, while staying at the same mass coordinate. Towards these smaller radii, $b=\ln[(T/T_0)^{5/2}/(P/P_0)]$ decreases (see also Fig.~\ref{Fig:absprofiles}) as the pressure increases more rapidly than the temperature towards the stellar core, while the mean molecular weight as well as $\delta\mu$ and thus $a$ remain constant at the $\mu$-disc. This decrease in $b$ at $r_{\mu \textrm{\tiny{-disc}}}$ is highly correlated with the decrease in $\delta s$ (see Fig.~\ref{Fig:bds}). 

Interpreting the correlation between $b=\ln[(T/T_0)^{5/2}/(P/P_0)]$ and $\delta s$ as a causality would mean that the decrease in $\delta s$ is caused by the $\mu$-disc moving to smaller radii, where $b$ is lower. By extension, given that the reduction in $\delta s$ is responsible for the reduction in $s_{\textrm{\tiny{surface}}}$ (Fig.~\ref{Fig:sderivs}), this would imply that the $\mu$-disc moving inwards in the contracting stellar core causes the onset of the bump. 

To support our interpretation of causality, we computed a stellar evolution track where we artificially induced a small increase ($0.01\%$) in temperature throughout the star at every time step to induce a change in $b$. For the pressure to be as unaltered as possible, we induced the same percental decrease in density across the model at every time step. For this alternative track we indeed found a change in $b$. While we also found a different morphology of the bump, the correlations between the luminosity and the specific entropy at the surface as well as between $\delta s$ and $b$ remained high similarly to those shown in Figs~\ref{Fig:HRD} \& \ref{Fig:bds}. Although this provides no definite proof of causality, it does support the interpretation of causality by showing that the change in the temperature, and with that the change in $b$, impacts the specific entropy as expected from Eq.~\ref{Eq:s1}.

\section{Results}
\begin{figure}
\centering
\includegraphics[width=\columnwidth, clip]{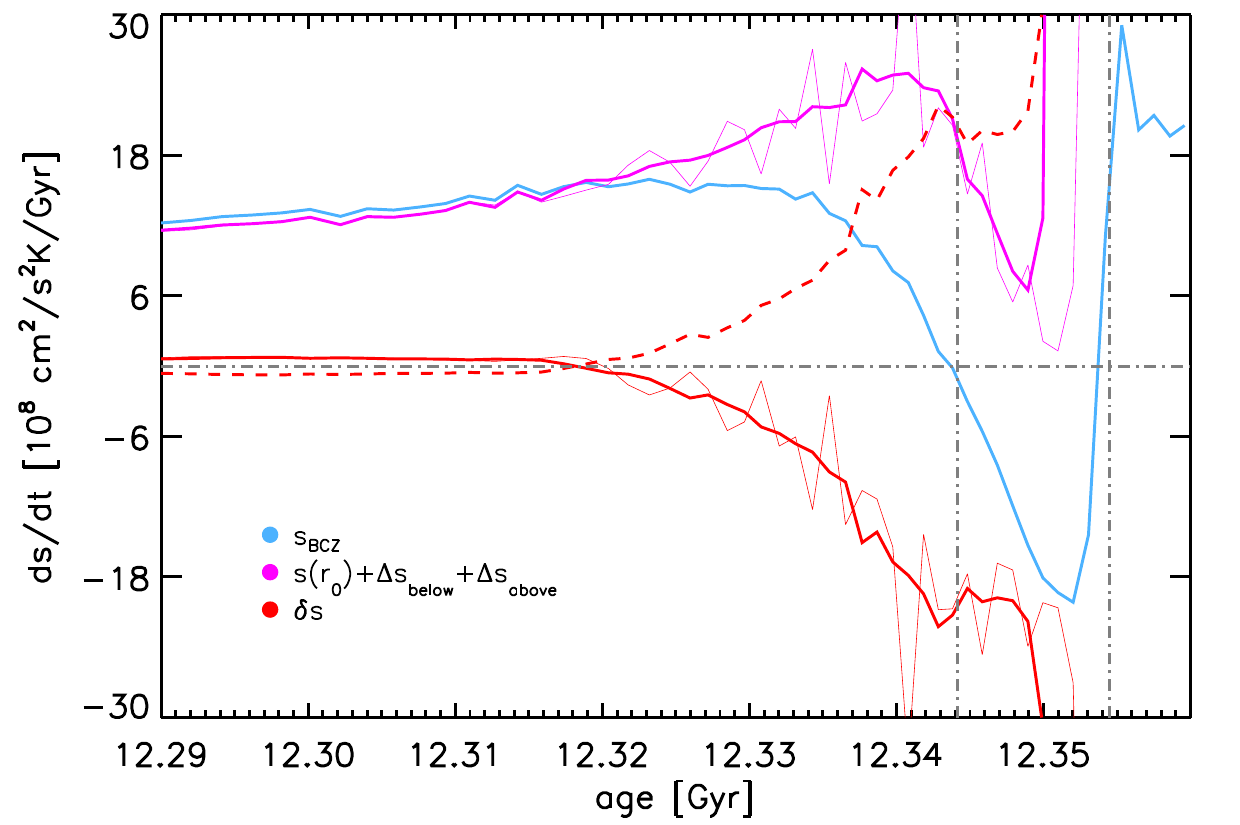}
\caption{The time derivatives of the specific entropy at the BCZ (light blue), in the radiative region (magenta) and at the $\mu$-disc (red) as a function of evolution. The thick lines are smoothed versions (boxcar with a width of seven points) to reduce the numerical noise of the derivatives. The red dashed line shows the derivative of $\delta s$ multiplied by $-1$. The horizontal grey dashed-dotted line indicates zero, i.e. no change over time. The vertical grey dashed-dotted lines indicate the time of luminosity maximum and minimum. We note that at the end point of the bump when the discontinuity vanishes, the monotonous increase of $s$ in the radiative layers returns everything `back to normal'.\label{Fig:sderivs}}
\end{figure}

With the analysis above, we show that the decrease in $\delta s$ provides a viable description for the onset of the luminosity bump. In the radiative layers around the $\mu$-disc the specific entropy increases with radius and time as expected (see the magenta curves in Figs~\ref{Fig:sparts} \& \ref{Fig:sderivs} for the actual values and derivatives). However, the contribution of the discontinuity, i.e. $\delta s$, decreases when the discontinuity moves to smaller radii following the receding of the BCZ. The decrease in $\delta s$ gradually slows down the expansion of the star and thus the increase in the luminosity. This gradual change is also reflected in the relatively smooth onset of the RGBB compared to the much sharper feature at the end of the bump.

The value of $\delta s$ reaches a critical value, when the decrease in $\delta s$ from one timestep to the next is equal to the increase in the specific entropy in the surrounding radiative layers during that timestep i.e. when
\begin{equation}
    -\frac{\partial \delta s}{\partial t} = \frac{\partial s_{\rm rad}}{\partial t}
\label{eq:dscrit}
\end{equation}
with $s_{\rm rad} = s(r_0)+ \Delta s_{\textrm{\tiny{below}}} + \Delta s_{\textrm{\tiny{above}}}$ (see also the point where the red-dashed and magenta lines cross in Fig.~\ref{Fig:sderivs}). Eq.~\ref{eq:dscrit} is equivalent to $\partial s_{\textrm{\tiny{BCZ}}}/\partial t =0$. From this time onwards
the specific entropy at the BCZ (and thus the stellar surface) starts decreasing and thus the star starts contracting. This transition from expanding to contracting marks the luminosity maximum of the bump. We note that the specific entropy in the radiative regions surrounding the discontinuity is continuously increasing (magenta line in Fig.~\ref{Fig:sderivs} is always positive), confirming that without the discontinuity there would not be a bump. Finally, we analysed stellar evolution tracks with different masses [1.0, 1.5, 2.0]~M$_{\odot}$ at solar metallicity and tracks with [Fe/H] = [$-$0.3, 0.0, $+$0.2] dex at solar mass. All tracks show qualitatively the same behaviour, indicating that the results presented here for a stellar evolution track with solar mass and metallicity are applicable across a range of mass and metallicity.

\section{Discussion and Conclusions}
We conclude here that the evolution of the difference in specific entropy $\delta s$ at the $\mu$-disc provides a viable description for the onset of the red-giant bump. The difference in $\delta s$ is predominantly determined by the underlying temperature ($T$) and pressure ($P$) at the location of the $\mu$-disc, which change when the $\mu$-disc moves to deeper layers. We derived this by recasting the long-standing open question regarding the mechanism behind the onset of the bump into specific entropy. If the specific entropy view is indeed responsible for the onset of the bump, this view shows for the first time that the movement of the $\mu$-disc to smaller radii and the different hydrostatic circumstances there are crucial. This would allow us to identify which changes in hydrostatics are responsible for the bump. While, \citet[e.g.][]{jcd2015} were able to make the general statement that hydrostatic changes at the $\mu$-disc are responsible for the onset of the bump, a description in terms of specific entropy will allow us to make a significant step forward by identifying more precisely which changes in the thermal structure are relevant.

This viable description of the mechanism behind the onset of the bump provides crucial and new insights for the exploitation of the bump as an anchor point in stellar evolution theory. Such added knowledge of the onset of the bump, could potentially be used to make more informed choices for changes to the physics in stellar models that go beyond calibration of the overshoot, to place the bump in stellar evolution models at the correct luminosity while at the same time aligning them with measured surface abundances. For example, as we found that the onset of the bump likely depends on the local temperature and pressure at the $\mu$-disc, we speculate that it may be necessary to look at the equation of state. Such investigation is considered to be beyond the scope of the current work.

\begin{acknowledgments}
SH thanks the following people for very valuable discussions, comments and insights: Sarbani Basu, Derek Buzasi, Lucas Eekhof, Yvonne Elsworth, Tobias van Lier, Jonas M\"uller, Tamara Rogers, and all members of the TOS group at HITS. \textbf{SH also thanks the anonymous referee for constructive feedback that improved the manuscript significantly.} We acknowledge funding from the ERC Consolidator Grant `DipolarSound' (grant agreement \# 101000296) and the Klaus Tschira Foundation.
\end{acknowledgments}


\bibliography{references}{}
\bibliographystyle{aasjournalv7}



\end{document}